\documentclass[twocolumn]{aastex61}

\usepackage{epstopdf}
\usepackage{amssymb}
\usepackage{graphicx}
\usepackage{natbib}

\newcommand{\cha}{$^{12}$CH$^+$}
\newcommand{\chb}{$^{13}$CH$^+$}
\newcommand{\cda}{$^{12}$CD$^+$}
\newcommand{\cdb}{$^{13}$CD$^+$}

\newcommand{\ch}{CH$^+$}

\submitjournal{the Astrophysical Journal on February 1, 2018.  Accepted on February 26, 2018}

\shorttitle{spectra of CH$^+$}
\shortauthors{Dom\'{e}nech et al.}
\begin{document}

\title{First laboratory detection of vibration-rotation transitions of \cha\ and \chb\
and improved measurement of their rotational transition frequencies}

\author[0000-0001-8629-2566]{Jos\'e L. Dom\'enech}
\affiliation{Instituto de Estructura de la Materia (IEM-CSIC), Serrano 123, E28006 Madrid, Spain}
\author{Pavol Jusko}
\altaffiliation{current affiliation: Institut de Recherche en Astrophysique et Plan\'etologie (IRAP), Universit\'e 
de Toulouse (UPS), CNRS, CNES, 
9 Av. du Colonel Roche, 31028 Toulouse Cedex 4, France}
\affiliation{I. Physikalisches Institut, Universit\"at zu K\"oln, Z\"ulpicher Str.~77,
50937 K\"oln, Germany}
\author[0000-0002-1421-7281]{Stephan Schlemmer}
\affiliation{I. Physikalisches Institut, Universit\"at zu K\"oln, Z\"ulpicher Str.~77,
50937 K\"oln, Germany}
\author[0000-0003-2995-0803]{Oskar Asvany}
\affiliation{I. Physikalisches Institut, Universit\"at zu K\"oln, Z\"ulpicher Str.~77,
50937 K\"oln, Germany}
\correspondingauthor{Oskar Asvany}
\email{asvany@ph1.uni-koeln.de}
\date{\today}

\begin{abstract}
The long-searched C-H stretches of the fundamental ions  CH$^+$ and $^{13}$CH$^+$  have been observed for the first time
in the laboratory.
For this, the state-dependent attachment of He atoms to these ions at cryogenic temperatures has been exploited 
to obtain high-resolution rovibrational data. In addition, the lowest rotational transitions of CH$^+$, $^{13}$CH$^+$ and CD$^+$
have been revisited and their rest frequency values improved substantially.

\end{abstract}



\section{Introduction}
Methylidynium, \ch, was the first molecular ion and one of the first interstellar molecules observed in space, 
through its absorption lines in the visible region \citep{Dunham1937}.  It was identified in the laboratory by \citet{Douglas1941}. Since then, it has proven to be almost ubiquitous, 
and it has been observed in different environments through its electronic transitions, both in emission (see e.g. \citet{Hobbs2004}) and 
absorption (see e.g. \citet{Weselak2008}). 
The observation of the \ch ~ $J=1-0$ 
rotational transition at $\sim835$ GHz from ground, however, is hindered by a strong telluric oxygen line. 
One way out is the observation of the \chb\ isotopologue at $\sim830$ GHz, 
which has been detected with the CSO and APEX telescopes under very good weather conditions \citep{fal05,men11}, 
or the observation of \ch\ in red-shifted extragalactic sources
in the mm-wave region, as detected  with ALMA towards the blazar PKS 1830-211 \citep{Muller2017} and a group of starburst galaxies \citep{fal17}. 
Local observations of the rotational lines of \ch\ require satellite missions.
\citet{cer97} observed pure rotational transitions in emission with the ISO satellite, and {\it Herschel} has provided 
extensive data both in absorption against bright sub-mm sources (e.g. \citet{Falgarone2010a,Godard2012}) 
and in emission in warm gas environments (e.g. \citet{nag13,Parikka2017}).


The astropysical relevance of \ch\ stems from its role as initiator and key molecule in the carbon chemistry in the 
interstellar medium, and from its, yet to be understood, high abundance in diffuse clouds.  This is a problem that keeps puzzling 
astronomers since the 70's \citep{Dalgarno1976}, because the observed abundances in diffuse clouds are well above the values 
predicted by even the most recent models (e.g. \citet{Valdivia2017} and references therein).  The main formation reaction is 
believed to be the reaction ${\rm C^+ + H_2\rightarrow CH^+ + H}$, which is endothermic (by 0.37 eV, or $\sim$4300 K) when H$_2$ 
is in its ground vibrational state \citep{Hierl1997}.    Therefore, in order to surmount the formation barrier in diffuse clouds, 
different sources of suprathermal energy have been suggested, like C-shocks \citep{Flower1998}, magneto-hydrodynamics shocks \citep{Lesaffre2013} 
or turbulent dissipation \citep{Godard2014}.  These processes could favor the formation of \ch\ and overcome the efficient destruction mechanisms of \ch, 
mainly reactions with  H and H$_2$ and recombination with electrons.  
In warmer environments like PDRs in dense clouds, the vibrational excitation of H$_2$ makes the formation reaction above exothermic, 
proceeding at almost the Langevin rate \citep{Agundez2010,Zanchet2013}.

Despite its astrophysical importance, accurate laboratory data on its rotational spectrum did not become available until very recently \citep{ama10b,yu15b}.  
Most previously available information came from the electronic emission spectrum of the $A^1\Pi−-X^1\Sigma$ system, the most notable experiments being 
those of \citet{Carrington1982} and \citet{Hakalla2006} on \cha, and those of \citet{Bembenek1987,Bembenek1997,Bembenek1997a} on \chb, \cda\ and \cdb, respectively.
The astronomical observation by \citet{cer97} was of too low resolution for an accurate determination of the lowest frequency transitions, 
and a previous laboratory measurement of the $J=1-0$ transition of \cha ~\citep{Pearson2006} was proven to be too low by 58~MHz.
\citet{ama10b}  finally provided  accurate frequencies for that transition, as well as  for \chb\ and \cda.
In contrast, the vibration-rotation IR spectrum had eluded its observation until now.
Previous to the work described here, the rovibrational lines of \cha\ were also searched for with a difference frequency spectrometer 
coupled to a hollow cathode discharge \citep{Cueto2014,Domenech2016} without success, thus emphasizing the elusive 
character of the vibration-rotation spectrum of this fundamental ion, already pointed out by \citet{Rehfuss1992}.  An exhaustive compilation of the spectroscopic laboratory literature, used in a direct potential fit of all available frequency data, can be found in  \citet{Cho2016}.

Regarding transition intensities, \citet{Follmeg1987} calculated the permanent electrical dipole moment of \ch\ to be $\mu_0=1.679$~D, and 
the transition dipole moment for the $v=1-0$ band $\mu_{1-0}=0.016$~D.   These values are in reasonable agreement with those derived 
from \citet{Sauer2013} of  1.7~D and 0.0196~D, respectively.  \citet{Cheng2007} calculated $\mu_0=1.683$~D, 
but they did not derive a vibrational transition moment.
The low value of $\mu_{1-0}$, together with the high reactivity of \ch, which is rapidly destroyed by 
collisions with H$_2$, O$_2$ or H$_2$O typically present in discharge tubes as traces, 
could explain the non-observation in the cold cathode experiments.

In this work, we present the first direct observation of vibration-rotation lines of \cha\ and \chb, with accuracies 
better than 1~MHz, as well as significantly improved measurements of the $J=1-0$ rotational transitions already available.


\section{Experimental setup}

\begin{figure}
 \includegraphics[width=\columnwidth]{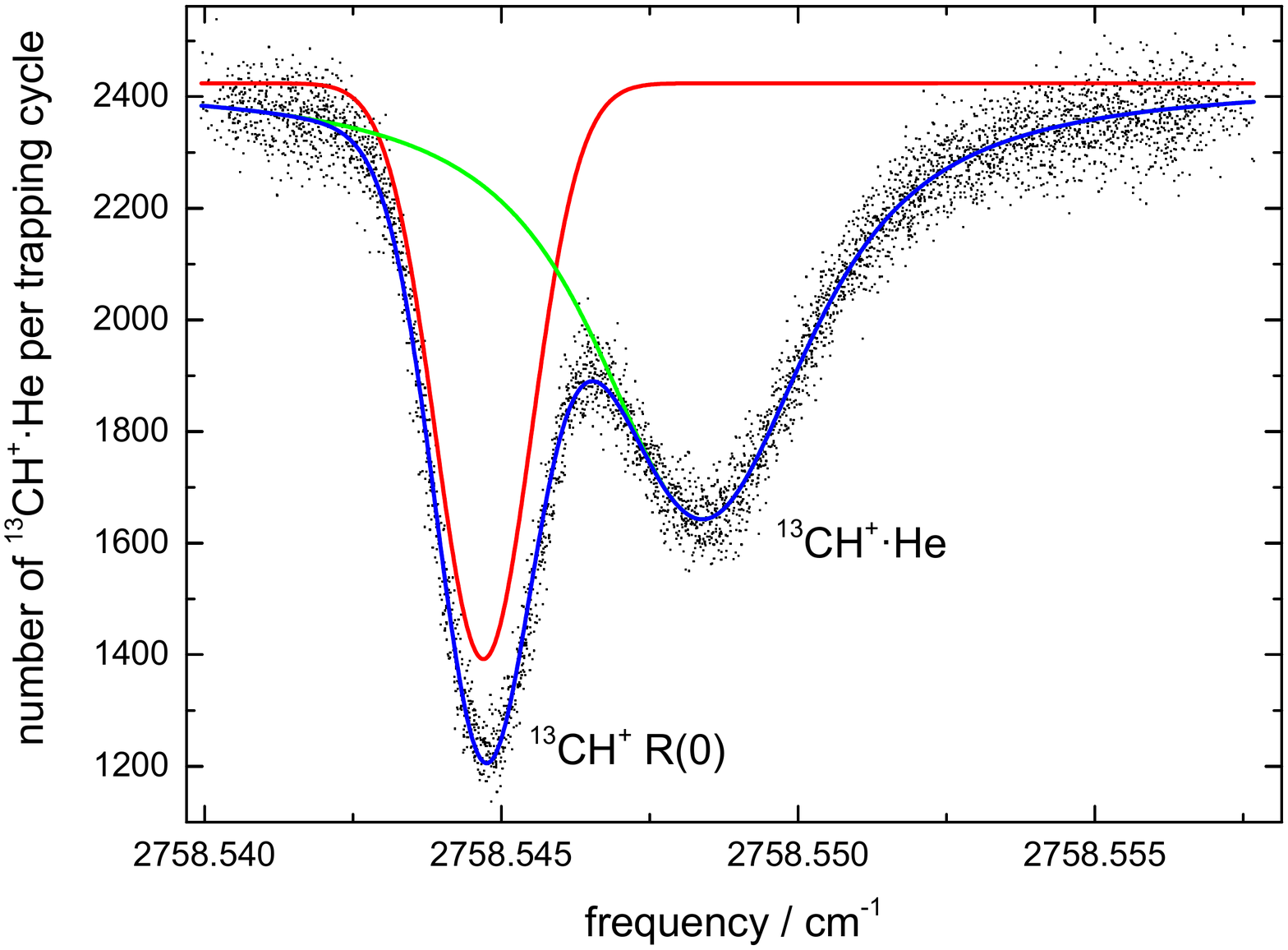}
 \caption{\label{fig1} Accidental blending of the R(0) transition of $^{13}$CH$^+$ with one unassigned transition of  $^{13}$CH$^+$-He.
We observe the state-dependent He-attachment to $^{13}$CH$^+$  and the predissociation line of $^{13}$CH$^+$-He 
simultaneously by counting the  $^{13}$CH$^+$-He ions (mass 18~u) after the 700~ms long trapping time during which the ions are irradiated. 
These two species can be distinguished by their spectral position and their lineshape.
The shown  $^{13}$CH$^+$ line is Doppler broadened and has a Gaussian shape (red), giving a temperature $T=13$~K, while the  $^{13}$CH$^+$-He
line is additionally lifetime broadened, having a Voigt profile (green). Its Lorentzian contribution yields a lifetime $\tau = 1.6$~ns
of the vibrationally excited state.}
\end{figure}

The spectroscopy of CH$^+$ was facilitated by a cryogenic ion trap experiment, in which the mentioned 
parasitic reactants are frozen out.
As the applied trapping setup (\cite{asv10,asv14}) and the  
action schemes for rovibrational  as well as pure rotational
spectroscopy have been  thoroughly documented by \cite{asv14,sav15,jus16,jus17,dom17,sto16,bru14,bru17}, 
only a brief description is given here.   
The CH$^+$ (similarly  $^{13}$CH$^+$ and CD$^+$) ions have been generated in a storage ion source by bombarding the precursor gas
(CH$_4$ Linde 5.5, $^{13}$CH$_4$ Sigma Aldrich 99\%, or CD$_4$ Cambridge Isotope Laboratories 99\%, respectively) with electrons (with energies in the range 30-40~eV).
A pulse of about tenthousand mass-selected ions was injected into 
the 22-pole ion trap filled with about 10$^{14}$ cm$^{-3}$ He.
During the trapping time of  700~ms, the complexes CH$^+$-(He)$_n$ ($n=1-4$) formed 
by three-body collisions. The detection of the resonant absorption of the continuous radiation (submm or IR)
by the naked ion was achieved by observing 
the decrease of complexes with $n=1$ (masses 17~u or 18~u, respectively).   
The IR radiation was provided by an Aculight Argos Model 2400 cw OPO (optical parametric oscillator), 
with the power in the trap reaching hundreds of mW. The frequency of the IR radiation has been measured
by our frequency comb system (\cite{asv12}) with an accuracy in the range of 80~kHz. 
The submm-wave  radiation was supplied by a synthesizer driving a  multiplier chain source
(Virginia Diodes, Inc.) covering the range 80-1100~GHz.
The nominal submm-wave  power at the frequency of the fundamental rotational transition of CH$^+$ was about 6~$\mu$W. 
Both the used synthesizer (Agilent E8257D) as well as the frequency comb system were locked to  
a rubidium atomic clock (with a typical stability of 10$^{-11}$).

   
\section{Rovibrational and  rotational  transitions}

\begin{figure}
 \includegraphics[width=.5\textwidth]{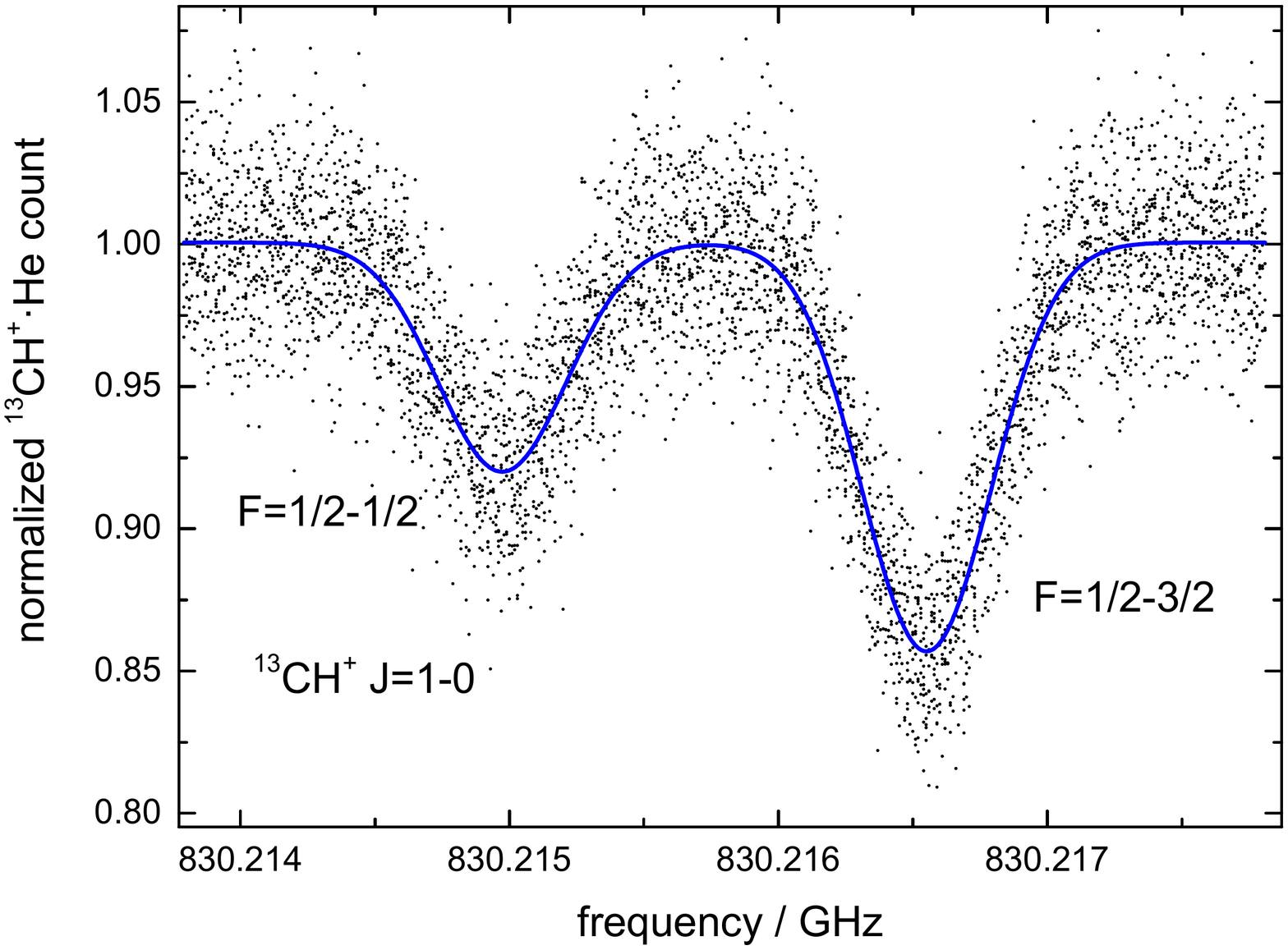}
 \caption{\label{fig2}  The $J = 1 \leftarrow 0$ rotational transition of  $^{13}$CH$^+$
   recorded at a nominal temperature of 4~K,
   showing two clearly resolved hyperfine components. 
 Every ion count (black dot) has been normalized by
a subsequent count with the submm-wave source being off-resonant.
The hyperfine structure is caused by the $I(^{13}\rm{C})=1/2$ nuclear spin
   of the  $^{13}$C nucleus, whereas the influence of the hydrogen nuclear spin  $I(^{13}\rm{H})=1/2$ is 
negligible here: $F = J + I(^{13}\rm{C})$.}
\end{figure}

The low temperature requirement of our spectroscopy method (with a a maximum operation temperature 
of about 20~K for the CH$^+$-He case) together with the large rotational constants of these light hydrides
permitted to investigate only a few low-lying transitions in this work (albeit with very high precision).
We based our search on the available accurate predictions for the $^1\Sigma$ ground state provided by \cite{mul10}, \cite{yu16b} and the 
CDMS database (\cite{CDMS_3}). 
For the isotopologoues   CH$^+$ and $^{13}$CH$^+$, four rovibrational and one rotational transition have been found and recorded,
which are summarized in Tables~\ref{tab1} and \ref{tab2}.
Whereas the IR transitions are detected for the first time in this work, the pure rotational lines
have been measured before by \cite{ama10b} and \cite{yu15b} in a discharge cell  and their values are also included in the Tables.  
Interestingly, in our IR scans we also observed rovibrational transitions of the complex CH$^+$-He (and similarly  $^{13}$CH$^+$-He)
which are detected by their photodissociation upon absorption.
These lines can be distinguished
by their lifetime-broadened Lorentzian shape and their spectral position, see Fig.~\ref{fig1}.
This is very similar to the observations made for the H$_3^+$-He system by \cite{sav15},
for which both the naked ion and its He complex have been observed in the same spectrum.
The spectrum of CH$^+$-He will be analyzed in a future publication. \\
An example for a pure rotational measurement is shown in Fig.~\ref{fig2}. We choose to show
the  $J = 1 \leftarrow 0$  transition of $^{13}$CH$^+$,  as this transition can be observed from ground (\cite{fal05,men11}),
and because it exhibits hyperfine structure first seen by \cite{ama10b,ama10c} in the laboratory.
The hyperfine splitting is clearly resolved in our
cryogenic experiment with a 1:2 intensity ratio. For this rotational measurement as well as for the rovibrational measurements
of the CH$^+$ isotopologues, we generally detect pure Gaussian profiles 
with a measured Doppler temperature close to $T=12$~K (with the ion trap at nominal 4~K). Additional power broadening, as sometimes observed in
trap-based mm-wave spectroscopy (see e.g.\  \cite{toe16,bru17}) is not observed here due to the limited power applied.

The fundamental rovibrational transitions of CD$^+$ were not accessible by our OPO system, 
but we revisited the two lowest rotational transitions which are listed  
in Table~\ref{tab3}. We could improve the value for the  $J = 2 \leftarrow 1$ transition by nearly 2 orders of magnitude,
while the accurate value for $J = 1 \leftarrow 0$ given very recently by \cite{bru17} 
(measured in the Cologne laboratories in a different trap setup)  is confirmed.

\begin{table}[h]  
 \caption{\label{tab1} Measured frequencies of rotational and  
 rovibrational transitions of  CH$^+$, and comparison to former work. 
The accuracy of our measurements is 
 given by locking the frequency comb and the mm-wave synthesizer to a rubidium clock.
 The narrow Doppler widths lead to relative precisions close to 1~ppb. Every line in this table has been measured at least
 five times, with the final error given in parentheses.}
 \begin{tabular}{clr@{}ll}
   $(v,J)  \leftarrow  (v,J)$ & this work       & \multicolumn{2}{c}{former work}            & unit\\
\hline
    $(0,1)  \leftarrow  (0,0)$  & 835137.4408(10)   &   835137&.504(20) $^a$       & MHz\\
    $(0,2)  \leftarrow  (0,1)$  &                   &  1669281&.361(100)$^b$      & MHz\\
    $(0,3)  \leftarrow  (0,2)$  &                   &  2501443&.102(100)$^b$      & MHz\\[.15cm]

   $(1,0)  \leftarrow  (0,1)$   &  2711.812948(22)  &        &          & cm$^{-1}$  \\
   $(1,1)  \leftarrow  (0,0)$   &  2766.548226(6)   &        &          & cm$^{-1}$   \\
   $(1,2)  \leftarrow  (0,1)$   &  2792.414725(7)   &        &          & cm$^{-1}$  \\
   $(1,3)  \leftarrow  (0,2)$   &  2817.237730(65)  &        &          & cm$^{-1}$  \\
\hline
 \end{tabular}
$^a$ \cite{ama10b} \\
$^b$  \cite{yu15b}
\end{table}

\begin{table}[h] 
 \caption{\label{tab2} 
  Measured frequencies of rotational and rovibrational transitions of  $^{13}$CH$^+$, 
and comparison to former work. The two hyperfine components of the 
   $(0,1)  \leftarrow  (0,0)$ transition (due to the spin I=1/2 of the $^{13}$C nucleus) have been clearly 
   separated in our cryogenic measurements. Every line in this table has been measured at least
 five times, with the final error given in parentheses. }
  \begin{tabular}{clr@{}l@{ }l}
  \scriptsize
    $(v,J,F)  \leftarrow  (v,J,F)$  & this work       &  \multicolumn{2}{c}{former work}  & unit\\
\hline
    $(0,1,\frac{1}{2})  \leftarrow  (0,0,\frac{1}{2})$  & 830214.9715(23) &  830215&.004(30)$^a$          & MHz\\
    $(0,1,\frac{3}{2})  \leftarrow  (0,0,\frac{1}{2})$  & 830216.5505(10) &  830216&.640(30)$^a$         & MHz\\
    $(0,2)  \leftarrow  (0,1)$  &                         &  1659450&.286(100)$^b$      & MHz\\
    $(0,3)  \leftarrow  (0,2)$  &                         &  2486725&.636(100)$^b$      & MHz\\[.15cm]

   $(1,0)  \leftarrow  (0,1)$   &       2704.128959(27) &          &        & cm$^{-1}$  \\
   $(1,1)  \leftarrow  (0,0)$   &       2758.544719(3)  &          &        & cm$^{-1}$   \\
   $(1,2)  \leftarrow  (0,1)$   &       2784.264862(9)  &          &        & cm$^{-1}$  \\
   $(1,3)  \leftarrow  (0,2)$   &       2808.950977(38) &          &        & cm$^{-1}$  \\
\hline
 \end{tabular}
$^a$ \cite{ama10b} \\
$^b$  \cite{yu15b}
\end{table}

\begin{table}[h]  
 \caption{\label{tab3}  Measured frequencies of rotational transitions of  CD$^+$ 
with comparison to literature values. 
  The values  are derived from typically ten independent measurements.
  The final errors are given in parentheses.}
 \begin{tabular}{cr@{}lr@{}ll}
   $(v,J)  \leftarrow  (v,J)$ & \multicolumn{2}{c}{this work}  & \multicolumn{2}{c}{former work} & unit  \\   
\hline
    $(0,1)  \leftarrow  (0,0)$  &   453521&.8530(6)                &    453521&.8509(7)$^a$             & MHz  \\
                                &         &                        &    453521&.851(20)$^b$             & MHz  \\
    $(0,2)  \leftarrow  (0,1)$  &   906752&.1649(17)               &   906752&.156(100)$^c$              & MHz \\
    $(0,3)  \leftarrow  (0,2)$  &         &                        &  1359399&.836(100)$^c$             & MHz  \\
    $(0,4)  \leftarrow  (0,3)$  &         &                        &  1811173&.908(100)$^c$             & MHz  \\
\hline
 \end{tabular}
$^a$ \cite{bru17} \\
$^b$ \cite{ama10b} \\
$^c$  \cite{yu15b}
\end{table}

\section{Results and Outlook}
\label{spec_parameters}

The submm-wave frequencies given in Tables~\ref{tab1} through \ref{tab3} compare very favourably with other work, 
the deviations being less than 3~$\sigma$ compared to our accurate values.  
Also the predictions for the IR transitions given currently on the CDMS database (\cite{CDMS_3}) turned out to be very accurate, 
with less than 0.01~cm$^{-1}$ difference for the low-J \ch\ transitions given in Table~\ref{tab1}.
The frequencies available in Tables~\ref{tab1} and \ref{tab2} have been used to determine the spectroscopic parameters
in the ground and excited vibrational state. For CH$^+$, the known FIR transitions detected
with the ISO satellite by \cite{cer97} helped to constrain the distortion constants in the ground state.
For $^{13}$CH$^+$, the well-resolved hyperfine components allow to determine the 
spin-rotation coupling constant $C_I$ with high precision.
The obtained values can  be found in Tables~\ref{param1} and~\ref{param2}.

\begin{table}
\begin{center}
\caption{\label{param1} The best fit spectroscopic parameters of CH$^+$  
are obtained by fitting the data given in Table~\ref{tab1} with the program PGOPHER (\cite{wes17}). 
To constrain the distortion constant $H$ in the ground state, also the high-J observations of \cite{cer97} have been used.
For the vibrationally excited state, a scaled fixed value has been adopted for $H$.
The numbers in parentheses give the uncertainty of the last digits.}
\begin{tabular}{lr@{}lr@{}ll}
\hline
Parameter 	 & \multicolumn{2}{c}{$v=0$} &  \multicolumn{2}{c}{$v=1$}           & unit \\
\hline
  $\nu$          &        &                 &  \multicolumn{2}{c}{2739.670097(5)}   &   cm$^{-1}$  \\[.15cm]
  $B$            &  417651&.595(4)          &     402974&.48(8)                     &  MHz \\
  $D$  	         &      41&.451(2)          &         40&.741(9)                    &  MHz  \\
  $H$  	         &       0&.0067(2)          &          0&.0065                     &  MHz   \\
\hline
\end{tabular}
\end{center}
\end{table}

\begin{table}
\begin{center}
\caption{\label{param2} The best fit spectroscopic parameters of $^{13}$CH$^+$  
are obtained by fitting the data given in Table~\ref{tab2} with the program PGOPHER (\cite{wes17}). 
For the distortion constants $H$, scaled fixed values have been adopted.
The numbers in parentheses give the uncertainty of the last digits.
The hyperfine spin-rotation coupling constant $C_I$ has been 
calculated and measured previously by \cite{sau95} and \cite{ama10c}.}
\begin{tabular}{lr@{}lr@{}ll}
\hline
Parameter 	 & \multicolumn{2}{c}{$v=0$} &  \multicolumn{2}{c}{$v=1$}          & unit \\
\hline
  $\nu$          &        &                  &  \multicolumn{2}{c}{2731.822034(8)} &  cm$^{-1}$  \\[.15cm]
  $B$            &  415189&.908(1)          &     400643&.48(14)                   &  MHz  \\
  $D$  	         &      40&.9609(7)         &         40&.27(1)                    &  MHz  \\
  $H$  	         &       0&.0066            &           0&.0064                     &  MHz  \\[.15cm]
  $C_I$          &       1&.053(1)          &            &                         &  MHz   \\
\hline
\end{tabular}
\end{center}
\end{table}


While the improvement of the rotational frequencies for the CH$^+$ isotopologues into the 1~ppb regime
are only of limited impact for the interpretation of astronomical data, the measured rovibrational transitions open up
their potential observation in the MIR regime. The low-lying vibration-rotation transitions measured in this work are in the range 3.5 - 3.7~$\mu$m  
and are thus well accessible from the ground.  Using the calculated vibrational transition dipole moment, and typical column densities and excitation 
temperatures derived from observations in the visible or mm-wave, one can estimate
the peak absorptions that could be expected in an infrared observation.
For example, \citet{Falgarone2010a} derived a column density  of $N=1.7\times10^{14}$\ molec\ cm$^{-2}$, an excitation temperature of 3 K, 
and velocity spreads of 5~km~s$^{-1}$ for the absorptions of \ch \ towards the massive star forming region DR21. For those conditions, 
most of the population resides in $J=0$, and the strongest IR absorption would be that of the $R(0)$ line  at 2766.548226(6)~cm$^{-1}$ (3.6146~$\mu$m).  
With $\mu_{1-0}=0.016$~D, the integrated line intensity is calculated to be

\begin{equation}
\int^{+\infty}_{-\infty}\alpha(\tilde{\nu})d\tilde{\nu}=3.0\times10^{-19}\ \mathrm{cm^2cm^{-1}molec^{-1}}   \,\,\,.
\end{equation}

\noindent The observed peak absorption will be directly proportional to the column density $N$, and inversely proportional to the 
observed linewidth. Assuming  the linewidth to be limited by the cloud velocity spread, and a triangular lineshape, 
the FWHM  is on the order of  $5/299792$ $\times$  2766.5~cm$^{-1}$ = 0.046~cm$^{-1}$
(that would imply using a spectrograph with a resolving power better than  R$\sim60000$).
 The opacity at the line center (or, equivalently, the peak absorption) would then  be 
\begin{equation}
\alpha_{\mathrm{peak}}= \frac{ 3.0\times10^{-19} \times  1.7\times10^{14}\  \mathrm{cm^{-1}}}{ 0.046\ \mathrm{cm^{-1}}   }    = 1.1\times10^{-3}                          
\end{equation}
and the fractional transmission against a bright continuum background source is then
\(
\tau_{\mathrm{peak}}=e^{-\alpha_{\mathrm{peak}}}= 0.999.
\)
Therefore, a detection at the 3$\sigma$ level requires a signal to noise ratio in the continuum of $\sim2700$.
Although the high resolution spectrograph CRIRES at the VLT could obtain spectra with signal to noise ratios of a 
few hundreds routinely, and values in excess of 5000 have been reported \citep{Smoker},  it is clear that a potential detection of this line in the IR from ground will be quite challenging.  
Also instruments like ISHELL 
at Mauna-Kea or the upcoming CRIRES+ at the VLT (also METIS at ELT and GMTNIRS at the GMT in a not so distant future) could provide the necessary 
sensitivity to detect these lines.  The lines of sight would be those of diffuse clouds in front of bright IR sources, like the dust 
envelopes around massive star formation regions. Other possible environments like dense PDRs can have $\sim100$ times higher \ch\ column densities, 
although the higher excitation temperatures would decrease the intensity of individual lines because of increased partitioning of the population.  
At an excitation temperature of 500 K, the strongest absorption would be that of the R(3) line, with about one tenth of the total intensity.  
Depending on the velocity structure and spread of the cloud, a detectable absorption could also be feasible.
Given the low vibrational transition dipole moment, mm-wave and vis-UV observations provide much better sensitivity for the study of this molecule in space.  
Notwithstanding the difficulties, IR observations will always be complementary, since they can target different sources, and with higher spatial resolution than mm-wave ones.  
Moreover, mm-wave observations from ground are quite difficult for the $J=1-0$ line of \chb\ and impossible for \cha, 
making  IR observations a potential future tool to contribute to the understanding of the processes at work in the regions where this ion is formed.

%


\section*{Acknowledgments}
This work (including the research visit of JLD in K\"oln) 
has been  supported by the Deutsche
Forschungsgemeinschaft (DFG) via SFB 956 project B2
and the Ger\"atezentrum "Cologne Center for Terahertz Spectroscopy". 
JLD acknowledges partial support from the Spanish MINECO through grant
FIS2016-77726-C3-1-P and from the European Research Council through grant
agreement ERC-2013-SyG-610256-NANOCOSMOS.
The authors thank Sandra Br\"unken for discussions and 
Holger M\"uller and Volker Ossenkopf-Okada 
for reading the manuscript before publication.




\begin{thebibliography}{}
\expandafter\ifx\csname natexlab\endcsname\relax\def\natexlab#1{#1}\fi
\providecommand{\url}[1]{\href{#1}{#1}}

\bibitem[{Ag{\'{u}}ndez {et~al.}(2010)Ag{\'{u}}ndez, Goicoechea, Cernicharo,
  Faure, \& Roueff}]{Agundez2010}
Ag{\'{u}}ndez, M., Goicoechea, J.~R., Cernicharo, J., Faure, A., \& Roueff, E.
  2010, Astrophys. J., 713, 662

\bibitem[{Amano(2010{\natexlab{a}})}]{ama10b}
Amano, T. 2010{\natexlab{a}}, Astrophys. J. Lett., 716, L1

\bibitem[{Amano(2010{\natexlab{b}})}]{ama10c}
---. 2010{\natexlab{b}}, The Journal of Chemical Physics, 133, 244305

\bibitem[{Asvany {et~al.}(2010)Asvany, Bielau, Moratschke, Krause, \&
  Schlemmer}]{asv10}
Asvany, O., Bielau, F., Moratschke, D., Krause, J., \& Schlemmer, S. 2010, Rev.
  Sci. Instr., 81, 076102

\bibitem[{Asvany {et~al.}(2014)Asvany, Br\"unken, Kluge, \& Schlemmer}]{asv14}
Asvany, O., Br\"unken, S., Kluge, L., \& Schlemmer, S. 2014, Appl. Phys. B,
  114, 203

\bibitem[{Asvany {et~al.}(2012)Asvany, Krieg, \& Schlemmer}]{asv12}
Asvany, O., Krieg, J., \& Schlemmer, S. 2012, Rev. Sci. Instr., 83, 093110

\bibitem[{Bembenek(1997{\natexlab{a}})}]{Bembenek1997}
Bembenek, Z. 1997{\natexlab{a}}, J. Mol. Spectrosc., 181, 136

\bibitem[{Bembenek(1997{\natexlab{b}})}]{Bembenek1997a}
---. 1997{\natexlab{b}}, J. Mol. Spectrosc., 182, 439

\bibitem[{Bembenek {et~al.}(1987)Bembenek, Cisak, \& Kepa}]{Bembenek1987}
Bembenek, Z., Cisak, H., \& Kepa, R. 1987, J. Phys. B At. Mol. Phys., 20, 6197

\bibitem[{Br\"unken {et~al.}(2014)Br\"unken, Kluge, Stoffels, Asvany, \&
  Schlemmer}]{bru14}
Br\"unken, S., Kluge, L., Stoffels, A., Asvany, O., \& Schlemmer, S. 2014,
  Astrophys. J. Lett., 783, L4

\bibitem[{Br\"unken {et~al.}(2017)Br\"unken, Kluge, Stoffels, Pérez-Ríos, \&
  Schlemmer}]{bru17}
Br\"unken, S., Kluge, L., Stoffels, A., Pérez-Ríos, J., \& Schlemmer, S.
  2017, Journal of Molecular Spectroscopy, 332, 67

\bibitem[{Carrington \& Ramsay(1982)}]{Carrington1982}
Carrington, A., \& Ramsay, D. 1982, Phys. Scr., 25, 272

\bibitem[{{Cernicharo} {et~al.}(1997){Cernicharo}, {Liu},
  {Gonz{\'a}lez-Alfonso}, {Cox}, {Barlow}, {Lim}, \& {Swinyard}}]{cer97}
{Cernicharo}, J., {Liu}, X.-W., {Gonz{\'a}lez-Alfonso}, E., {et~al.} 1997,
  Astrophys. J. Lett., 483, L65

\bibitem[{Cheng {et~al.}(2007)Cheng, Brown, Rosmus, Linguerri, Komiha, \&
  Myers}]{Cheng2007}
Cheng, M., Brown, J., Rosmus, P., {et~al.} 2007, Phys. Rev. A, 75, 012502

\bibitem[{Cho \& {Le Roy}(2016)}]{Cho2016}
Cho, Y.-S., \& {Le Roy}, R.~J. 2016, J. Chem. Phys., 144, 024311

\bibitem[{Cueto {et~al.}(2014)Cueto, Cernicharo, Barlow, Swinyard, Herrero,
  Tanarro, \& Dom{\'{e}}nech}]{Cueto2014}
Cueto, M., Cernicharo, J., Barlow, M.~J., {et~al.} 2014, Astrophys. J., 783, L5

\bibitem[{Dalgarno(1976)}]{Dalgarno1976}
Dalgarno, A. 1976, in At. Process. Appl. (Elsevier), 109--132

\bibitem[{Dom{\'{e}}nech {et~al.}(2016)Dom{\'{e}}nech, Drouin, Cernicharo,
  Herrero, \& Tanarro}]{Domenech2016}
Dom{\'{e}}nech, J.~L., Drouin, B.~J., Cernicharo, J., Herrero, V.~J., \&
  Tanarro, I. 2016, Astrophys. J., 833, L32

\bibitem[{Dom\'enech {et~al.}(2017)Dom\'enech, Schlemmer, \& Asvany}]{dom17}
Dom\'enech, J.~L., Schlemmer, S., \& Asvany, O. 2017, Astrophys. J., 849, 60

\bibitem[{Douglas \& Herzberg(1941)}]{Douglas1941}
Douglas, A.~E., \& Herzberg, G. 1941, Astrophys. J., 94, 381

\bibitem[{Dunham(1937)}]{Dunham1937}
Dunham, T. 1937, Publ. Astron. Soc. Pacific, 49, 26

\bibitem[{{Endres} {et~al.}(2016){Endres}, {Schlemmer}, {Schilke}, {Stutzki},
  \& {M{\"u}ller}}]{CDMS_3}
{Endres}, C.~P., {Schlemmer}, S., {Schilke}, P., {Stutzki}, J., \&
  {M{\"u}ller}, H.~S.~P. 2016, J. Mol. Spectrosc., 327, 95

\bibitem[{{Falgarone} {et~al.}(2005){Falgarone}, {Phillips}, \&
  {Pearson}}]{fal05}
{Falgarone}, E., {Phillips}, T.~G., \& {Pearson}, J.~C. 2005, Astrophys. J.
  Lett., 634, L149

\bibitem[{Falgarone {et~al.}(2017)Falgarone, Zwaan, Godard, Bergin, Ivson, \&
  many more}]{fal17}
Falgarone, E., Zwaan, M., Godard, B., {et~al.} 2017, Nature, 548, 430

\bibitem[{Falgarone {et~al.}(2010)Falgarone, Ossenkopf, Gerin, Lesaffre,
  Godard, Pearson, Cabrit, Joblin, Benz, Boulanger, Fuente, G{\"{u}}sten,
  Harris, Klein, Kramer, Lord, Martin, Martin-Pintado, Neufeld, Phillips,
  R{\"{o}}llig, Simon, Stutzki, van~der Tak, Teyssier, Yorke, Erickson, Fich,
  Jellema, Marston, Risacher, Salez, \& Schm{\"{u}}lling}]{Falgarone2010a}
Falgarone, E., Ossenkopf, V., Gerin, M., {et~al.} 2010, Astron. Astrophys.,
  518, L118

\bibitem[{Flower \& {Pineau des Forets}(1998)}]{Flower1998}
Flower, D.~R., \& {Pineau des Forets}, G. 1998, Mon. Not. R. Astron. Soc., 297,
  1182

\bibitem[{Follmeg {et~al.}(1987)Follmeg, Rosmus, \& Werner}]{Follmeg1987}
Follmeg, B., Rosmus, P., \& Werner, H.-J. 1987, Chem. Phys. Lett., 136, 562

\bibitem[{Godard {et~al.}(2014)Godard, Falgarone, \& {Pineau des
  For{\^{e}}ts}}]{Godard2014}
Godard, B., Falgarone, E., \& {Pineau des For{\^{e}}ts}, G. 2014, Astron.
  Astrophys., 570, A27

\bibitem[{Godard {et~al.}(2012)Godard, Falgarone, Gerin, Lis, {De Luca}, Black,
  Goicoechea, Cernicharo, Neufeld, Menten, \& Emprechtinger}]{Godard2012}
Godard, B., Falgarone, E., Gerin, M., {et~al.} 2012, Astron. Astrophys., 540,
  A87

\bibitem[{Hakalla {et~al.}(2006)Hakalla, K{\c{e}}pa, Szajna, \&
  Zachwieja}]{Hakalla2006}
Hakalla, R., K{\c{e}}pa, R., Szajna, W., \& Zachwieja, M. 2006, Eur. Phys. J.
  D, 38, 481

\bibitem[{Hierl {et~al.}(1997)Hierl, Morris, \& Viggiano}]{Hierl1997}
Hierl, P.~M., Morris, R.~A., \& Viggiano, A.~A. 1997, J. Chem. Phys., 106,
  10145

\bibitem[{Hobbs {et~al.}(2004)Hobbs, Thorburn, Oka, Barentine, Snow, \&
  York}]{Hobbs2004}
Hobbs, L.~M., Thorburn, J.~A., Oka, T., {et~al.} 2004, Astrophys. J., 615, 947

\bibitem[{Jusko {et~al.}(2016)Jusko, Konietzko, Schlemmer, \& Asvany}]{jus16}
Jusko, P., Konietzko, C., Schlemmer, S., \& Asvany, O. 2016, J. Mol.
  Spectrosc., 319, 55

\bibitem[{Jusko {et~al.}(2017)Jusko, Stoffels, Thorwirth, Br\"unken, Schlemmer,
  \& Asvany}]{jus17}
Jusko, P., Stoffels, A., Thorwirth, S., {et~al.} 2017, J. Mol. Spectrosc., 332,
  59

\bibitem[{Lesaffre {et~al.}(2013)Lesaffre, {Pineau des For{\^{e}}ts}, Godard,
  Guillard, Boulanger, \& Falgarone}]{Lesaffre2013}
Lesaffre, P., {Pineau des For{\^{e}}ts}, G., Godard, B., {et~al.} 2013, Astron.
  Astrophys., 550, A106

\bibitem[{{Menten} {et~al.}(2011){Menten}, {Wyrowski}, {Belloche},
  {G{\"u}sten}, {Dedes}, \& {M{\"u}ller}}]{men11}
{Menten}, K.~M., {Wyrowski}, F., {Belloche}, A., {et~al.} 2011, Astron.
  Astrophys., 525, A77

\bibitem[{{M{\"u}ller}(2010)}]{mul10}
{M{\"u}ller}, H.~S.~P. 2010, Astron. Astrophys., 514, L6

\bibitem[{Muller {et~al.}(2017)Muller, M{\"{u}}ller, Black, G{\'{e}}rin,
  Combes, Curran, Falgarone, Gu{\'{e}}lin, Henkel, Mart{\'{i}}n, Menten,
  Roueff, Aalto, Beelen, Wiklind, \& Zwaan}]{Muller2017}
Muller, S., M{\"{u}}ller, H. S.~P., Black, J.~H., {et~al.} 2017, Astron.
  Astrophys., 606, A109

\bibitem[{Nagy {et~al.}(2013)Nagy, {Van der Tak, F. F. S.}, {Ossenkopf, V.},
  {Gerin, M.}, {Le Petit, F.}, {Le Bourlot, J.}, {Black, J. H.}, {Goicoechea,
  J. R.}, {Joblin, C.}, {Röllig, M.}, \& {Bergin, E. A.}}]{nag13}
Nagy, Z., {Van der Tak, F. F. S.}, {Ossenkopf, V.}, {et~al.} 2013, Astron.
  Astrophys., 550, A96

\bibitem[{Parikka {et~al.}(2017)Parikka, Habart, Bernard-Salas, Goicoechea,
  Abergel, Pilleri, Dartois, Joblin, Gerin, \& Godard}]{Parikka2017}
Parikka, A., Habart, E., Bernard-Salas, J., {et~al.} 2017, Astron. Astrophys.,
  599, A20

\bibitem[{Pearson \& Drouin(2006)}]{Pearson2006}
Pearson, J.~C., \& Drouin, B.~J. 2006, Astrophys. J., 647, L83

\bibitem[{Rehfuss {et~al.}(1992)Rehfuss, Jagod, Xu, \& Oka}]{Rehfuss1992}
Rehfuss, B.~D., Jagod, M.-F., Xu, L.-W., \& Oka, T. 1992, J. Mol. Spectrosc.,
  151, 59

\bibitem[{Sauer \& Paidarov\'a(1995)}]{sau95}
Sauer, S.~P., \& Paidarov\'a, I. 1995, Chemical Physics, 201, 405

\bibitem[{Sauer \& {\v{S}}pirko(2013)}]{Sauer2013}
Sauer, S. P.~A., \& {\v{S}}pirko, V. 2013, J. Chem. Phys., 138, 024315

\bibitem[{Savi\'{c} {et~al.}(2015)Savi\'{c}, Gerlich, Asvany, Jusko, \&
  Schlemmer}]{sav15}
Savi\'{c}, I., Gerlich, D., Asvany, O., Jusko, P., \& Schlemmer, S. 2015, Mol.
  Phys., 113, 2320

\bibitem[{Smoker(2017)}]{Smoker}
Smoker, J. 2017, in 2017 ESO Calibration Work. Second Gener. VLT instruments
  friends, Santiago (Chile)

\bibitem[{{Stoffels} {et~al.}(2016){Stoffels}, {Kluge}, {Schlemmer}, \&
  {Br{\"u}nken}}]{sto16}
{Stoffels}, A., {Kluge}, L., {Schlemmer}, S., \& {Br{\"u}nken}, S. 2016,
  Astron. Astrophys., 593, A56

\bibitem[{T\"opfer {et~al.}(2016)T\"opfer, Jusko, Schlemmer, \& Asvany}]{toe16}
T\"opfer, M., Jusko, P., Schlemmer, S., \& Asvany, O. 2016, Astron. Astrophys.,
  593, L11

\bibitem[{Valdivia {et~al.}(2017)Valdivia, Godard, Hennebelle, Gerin, Lesaffre,
  \& {Le Bourlot}}]{Valdivia2017}
Valdivia, V., Godard, B., Hennebelle, P., {et~al.} 2017, Astron. Astrophys.,
  600, A114

\bibitem[{Welty {et~al.}(2006)Welty, Federman, Gredel, Thorburn, \&
  Lambert}]{Welty2006}
Welty, D.~E., Federman, S.~R., Gredel, R., Thorburn, J.~A., \& Lambert, D.~L.
  2006, Astrophys. J. Suppl. Ser., 165, 138

\bibitem[{Weselak {et~al.}(2008)Weselak, Galazutdinov, Musaev, \&
  Kre{\l}owski}]{Weselak2008}
Weselak, T., Galazutdinov, G., Musaev, F., \& Kre{\l}owski, J. 2008, Astron.
  Astrophys., 479, 149

\bibitem[{Western(2017)}]{wes17}
Western, C.~M. 2017, Journal of Quantitative Spectroscopy and Radiative
  Transfer, 186, 221

\bibitem[{Yu {et~al.}(2015)Yu, Drouin, Pearson, \& Amano}]{yu15b}
Yu, S., Drouin, B.~J., Pearson, J.~C., \& Amano, T. 2015, in Contribution RD06,
  70th International Symposium on Molecular Spectroscopy

\bibitem[{Yu {et~al.}(2016)Yu, Drouin, Pearson, \& Amano}]{yu16b}
Yu, S., Drouin, B.~J., Pearson, J.~C., \& Amano, T. 2016, in Contribution MH01,
  71st International Symposium on Molecular Spectroscopy

\bibitem[{Zanchet {et~al.}(2013)Zanchet, Godard, Bulut, Roncero, Halvick, \&
  Cernicharo}]{Zanchet2013}
Zanchet, A., Godard, B., Bulut, N., {et~al.} 2013, Astrophys. J., 766, 80

\end{thebibliography}

\end{document}